\documentclass[amsmath,amssymb,aps,prx,twocolumn,groupedaddress,superscriptaddress,10pt,reprint,longbibliography]{revtex4-2}
\usepackage{graphicx}% Include Fig.~files
\usepackage{textcomp} %for text mu

\usepackage{exscale}

\usepackage{relsize}

\usepackage[colorlinks,linkcolor=blue,citecolor=blue,urlcolor=blue]{hyperref}

\begin{document}

\preprint{APS/123-QED}

\title{Enhanced Detection of Rotational Doppler Shift from Sunlight}% Force line breaks with \\

\author{Jiedong Yang}
\author{Yuan Li}
\author{Wuhong Zhang}
 \email{zhangwh@xmu.edu.cn}
\author{Lixiang Chen}
 \email{chenlx@xmu.edu.cn}
\affiliation{Department of Physics, Xiamen University, Xiamen 361005, China}

\date{\today}% It is always \today, today,

\begin{abstract}
The rotational Doppler effect, for which the frequency shift is proportional to the light's orbital angular momentum $\ell$ and the object's rotational speed ($\Delta f \propto\ell \Omega$), has proven to be a powerful tool for detecting the speed of rotational objects. However, the current detection technique is mainly based on coherent laser sources. There is scarce mention of using partially coherent light sources, let alone sunlight. In this work, we collect sunlight and direct it into the laboratory, where it is modulated into a partially coherent probing source and then realize rotational Doppler shift detection. Our study reveals that in low-light conditions, where background noise is stronger than the signal, the superposition of rotational Doppler signals at different wavelengths can significantly enhance the signal strength and improve the signal-to-noise ratio, enabling accurate measurement of the rotational speed of objects. Our research provides experimental validation for the application of sunlight in rotational Doppler shift detection, demonstrating its potential value for passive remote sensing.
\end{abstract}

\maketitle

\section{\label{sec:level1}Introduction}

With the increasing understanding of photonic orbital angular momentum (OAM), research and applications related to OAM have rapidly expanded \cite{r1,r2,r3}. 
%三篇引用的文章不对，要找最近的综述文章
In the field of remote sensing, the rotational Doppler effect (RDE) has garnered significant attention due to its ability to directly measure the angular velocity of rotating objects. The RDE occurs when an OAM-carrying light wave, whose Poynting vector forms an angle with the optical axis that varies with the radial coordinate during propagation, illuminates a rotating object along its axis of rotation, resulting in a frequency shift that enables the measurement of the object's angular velocity \cite{r4,r5,r6,r7,r8}. In 2013, the Padgett group achieved a breakthrough by successfully applying this effect to measure the rotational speed of an object \cite{r9}. This pioneering experiment marked a significant step forward in utilizing OAM for rotational measurements, opening new avenues for the development of OAM-based sensing technologies. Since the frequency shift of the RDE is proportional to the light's orbital angular momentum $\ell$ and the object's rotational speed  ($\Delta f \propto\ell \Omega$), using white light beams that are combined from different single-wavelength laser sources has successfully realized the rotational Doppler shift phenomenon \cite{r10}. The RDE in the microwave frequency regime \cite{r11}, in free space outside the laboratory environment \cite{r12} has further promoted the research area. Recently, a direct connection between the temporal evolution of high-order Pancharatnam-Berry phases in light and the rotational Doppler frequency shift was well established, showing a new way to engineer the frequency content of light \cite{r13}. The frequency-upconversion detection of the RDE was proposed to realize infrared probing of the angular velocity and symmetry of rotating objects, even at the photon count level \cite{r14}. Then, the nonlinear frequency conversion of RDE with third-harmonic generation was demonstrated to further extend the detected frequency bound \cite{r15}. A new RDE detection scheme using twisted photons generated in scattered fields was demonstrated to eliminate the requirement for high-purity vortex sources \cite{r16}. By using a spatiotemporal mapping technique, a single-shot rotational Doppler metrology was proposed for angular velocity measurement \cite{r17}.  More recently, some new ideas and interesting schemes have been proposed to expand the RDE research area, such as real-time measurement of noncoaxial rotational Doppler shifts \cite{r18}, symmetry-dependent Doppler shifts in waveguide systems \cite{r19}, spatially varying rotational Doppler shifts by using a single trapped ion \cite{r20}, angle-independent motion detection with acoustic RDE \cite{r21} and a new transceiver-integrated all-fiber rotational Doppler velocimetry technology \cite{r22}. Since the spiral phase content of OAM requires a coherent source, all the aforementioned RDE-related studies naturally opt for a coherent laser as the detection source. 

\begin{figure*}[htbp]
    \centering
    \includegraphics[width=1\linewidth]{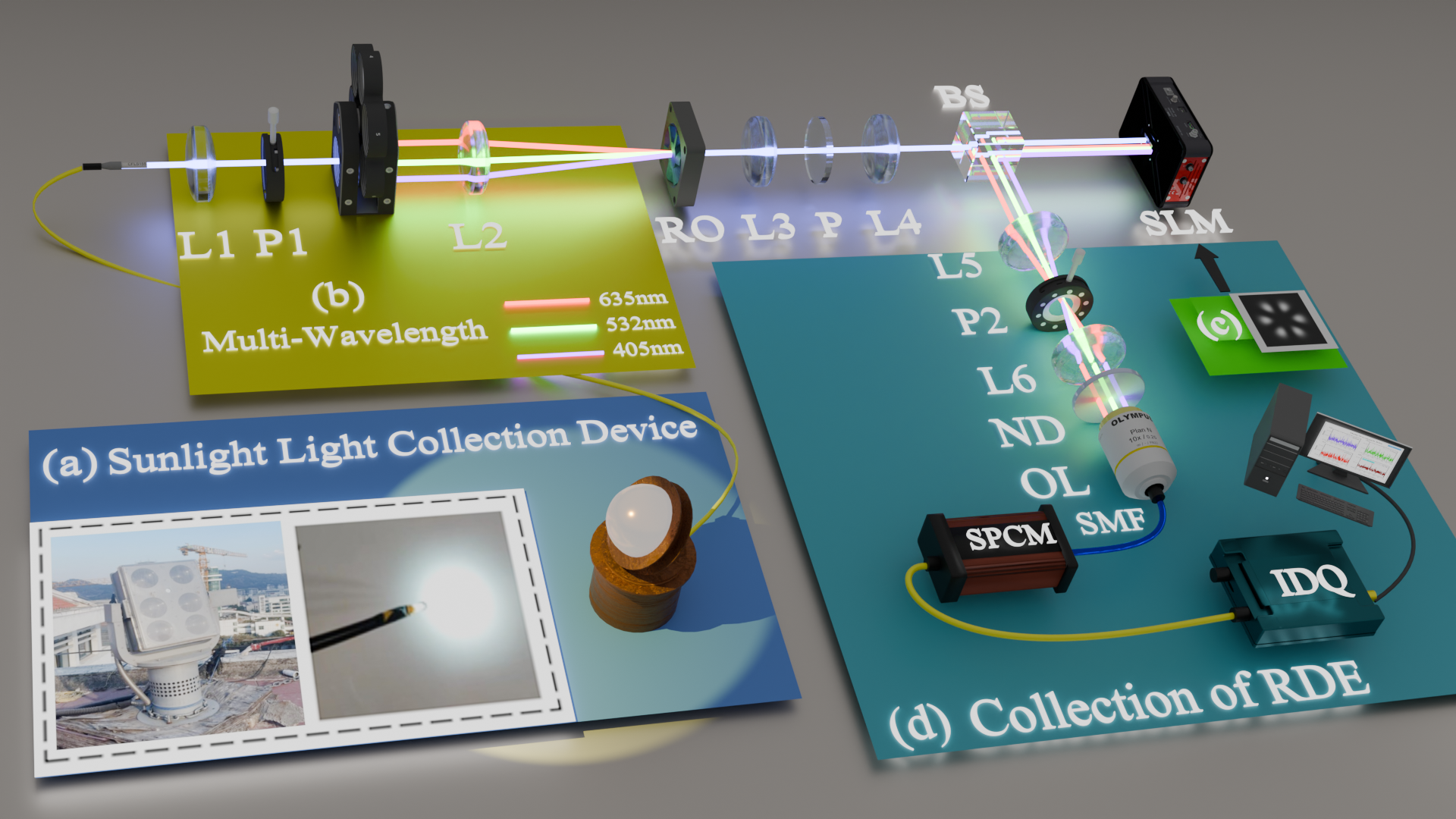}
    \caption{ Experimental setup for the rotational Doppler effect measurement using sunlight.The experiment utilizes sunlight as the light source to measure the rotational Doppler shift induced by a rotating object. (a) Sunlight collection system, (b) A series of bandpass filters filters, L:lens, RO:rotating object, P:polarizer, BS:beam splitter, ND：Neutral Density Filter, SLM:spatial light modulator, OL:Objective lens, SMF:single-mode fiber, SPCM:Single photon counting module, IDQ:ID Quantique.}
    \label{fig.1}
\end{figure*} 

However, it is noted that partially coherent light is also capable of carrying OAM \cite{r23,r24,r25,r26,r27,r28}. This characteristic opens up a promising research direction for the RDE based on partially coherent light. Some earlier attempts have been made to consider using partially coherent light to study the RDE. By using projector illumination, which can be seen as a spatially incoherent light source, Anderson et al. showed that an observable rotational Doppler shift signal can be generated \cite{r29}. By tracing the rotational motion of the coherent structure of the source, the angular velocity vectors (both magnitude and direction) of two anisotropic sub‐Rayleigh objects were successfully measured with ultrahigh precision \cite{r30}. They further demonstrated that such partially coherent light can effectively mitigate frequency broadening induced by both (a) beam-information loss due to occlusions and (b) beam misalignment in turbulent media \cite{r31}. Subsequently, building on this approach, off-axis angular velocity measurements were achieved using an LED-based spatially incoherent illumination source \cite{r32}. Collectively, these findings not only validate the feasibility of RDE sensing using partially coherent light but also offer experimental support for leveraging natural light in RDE detection. However, the illumination source in the above work only considers pseudothermal light sources or LED as the active lighting source. A truly incoherent light source, such as sunlight, which can be considered a passive lighting source, has not yet been studied. Considering sunlight as the illumination or detection source can make the RDE more practical for remote sensing by eliminating the need for a laser source, thus enhancing its real-world applicability and versatility. However, one of the core problems is that, under solar illumination, the number of photons reflected from rotating objects is typically insufficient for reliable signal detection, especially for long-distance sensing.

Here, we directly utilize sunlight as the passive lighting source  to measure the RDE. Besides, since the RDE is independent of the wavelength \cite{r10}, we take advantage of the broad spectral range of sunlight to achieve enhanced detection of the RDE under weak signals. We demonstrate that when the rotational Doppler signal of a single frequency is weak and submerged in complex background noise, one can retrieve the rotational Doppler signal by combining the Fourier transform spectral signals from multiple wavelengths, effectively suppressing background noise and improving the signal-to-noise ratio (SNR). Our study provides an enhanced detection method for measuring the angular velocity of rotating objects in a natural light environment and offers experimental validation for passive remote sensing based on natural light.

\section{Experimental setup}

\begin{figure}[htp]
    \centering
    \includegraphics[width=1\linewidth]{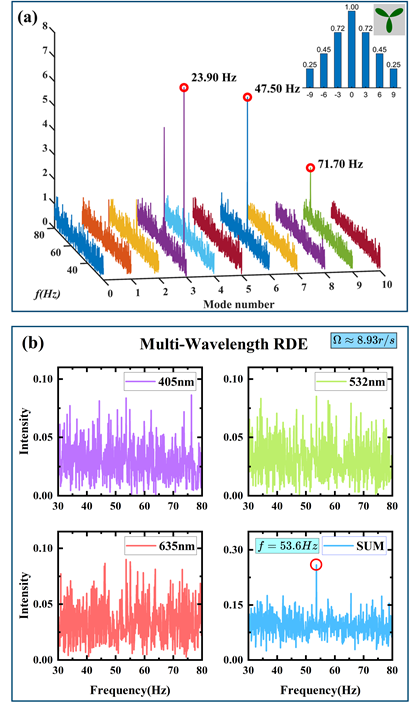}
    \caption{Rotational Doppler measurements under different conditions. (a) Frequency spectra for OAM superposition states $\ell = 0$ to $\ell = \pm10$ illuminating a threefold-symmetric rotating object ($\Omega = 4.00 \pm 0.01~\mathrm{r/s}$). Red markers indicate experimentally observed shift positions for symmetry-matched modes: $f_{3} = 23.90$ Hz, $f_{6} = 47.50$ Hz, and $f_{9} = 71.70$ Hz. The insets show a schematic of the three-leaf clover-shaped object and the pure OAM spectrum decomposition associated with the clover pattern. (b) $\Omega = 8.93 \pm 0.01~\mathrm{r/s}$ Experimental results of multi-wavelength superposition enhancement of rotational Doppler shift signal under sunlight.}
    \label{fig.2}
\end{figure}

As illustrated in Fig. \ref{fig.1}, we constructed an all-weather, automated sunlight-tracking system outdoors to collect sunlight and direct it to the lab for enhanced RDE tests. The sunlight collection apparatus operates similarly to an equatorial mount, which adjusts its altitude and azimuth automatically based on the current time and geographic coordinates. Additionally, a photosensitive chip is incorporated to enhance tracking accuracy. Once aligned, sunlight is directed vertically onto a Fresnel lens with a diameter of $10~\mathrm{cm}$ and then coupled into a $20~\mathrm{m}$-long, $2.5~\mathrm{mm}$-diameter plastic multimode fiber. The fiber delivers the sunlight into our optical laboratory, where it serves as the illumination source for the experiment, as shown in part (a) of Fig. \ref{fig.1}, The sunlight emitted from the fiber is collimated and expanded using a lens pair (L1 and L2) and a pinhole (P1). The pinhole can be used to modulate the spatial coherence of the sunlight, with smaller apertures yielding higher coherence. The expanded beam has a radius of approximately $13~\mathrm{mm}$, which matches the size of the rotating object. A set of  bandpass filters with central wavelengths of $405\pm 10~\mathrm{nm}$, $532\pm 10~\mathrm{nm}$, and $635\pm 10~\mathrm{nm}$ was used to selectively extract specific spectral components from the broadband solar spectrum to fulfill the demonstration of enhanced RDE detection with multi-wavelength, as illustrated in part (b). The collimated beam is then incident perpendicularly onto the rotating object (RO) which is a threefold-symmetric clover-shaped structure driven by a motor. The transmitted beam subsequently passes through a 4f imaging system comprising lenses L3 ($500~\mathrm{mm}$) and L4 ($100~\mathrm{mm}$) and is imaged onto a spatial light modulator (SLM). The SLM (FSLM-2K55P) features an active area of $12.29\times 6.91~\mathrm{mm}$ with a pixel size of $6.4~\mathrm{\mu m}$, supporting an operational wavelength range of $400-700~\mathrm{nm}$. For optimal modulation efficiency, the device requires horizontally polarized input light aligned with the SLM's horizontal orientation.
The polarizer (P) is employed to filter the horizontal polarization state of sunlight, thereby maximizing the diffraction efficiency of the SLM. 
The beam splitter (BS) redirects the diffracted light from the SLM for subsequent detection. 
According to our theoretical work \cite{r33}, the threefold rotational symmetry of a clover structure introduces distinctive features in the OAM spectrum of reflected/transmitted light. Specifically, the symmetry causes mode suppression—certain OAM modes exhibit zero amplitude while only selected modes (particularly $\ell =\pm3$) retain nonzero intensity. Specifically, the total power of the RDE signal under a partially coherent beam is:
\begin{equation}
\begin{split}
P(t)&=C\int_0^\infty \int_0^{2\pi}\sum_{\ell}I_\ell(\frac{r_1r_2}{\sigma^2})exp(\frac{-r^2}{\sigma^2})\\&[Re[A_{l+m}^*(r)A_{l-m}^*(r)exp(-2im\Omega t)]\\&+[Re[A_{l-m}^*(r)A_{l+m}^*(r)exp(2im\Omega t)]]rdrd\phi,
\label{eq7}
\end{split}
\end{equation}
where C is the amplitude of the cross-spectral density function in a Gaussian-Schell model beam, $I_\ell$ is the Modified Bessel Function of the First Kind, $\bm{r}_1,\bm{r}_2$ represents the position in the light field, and $\sigma$ is the coherence length of partially coherent light. Using this characteristic, we optimized the measurement scheme by selectively detecting only the surviving OAM components. To this end, we encoded a hologram corresponding to the $\ell =\pm3$ superposition state, which is shown in part (c), onto the SLM to modulate the input signal. Then, the reflected beam after the BS passes through a second 4f filtering system composed of lenses L5 ($100~\mathrm{mm}$), L6 ($100~\mathrm{mm}$), and a pinhole (P2) to filter out the desired first-order diffracted component.
The first-order diffracted filtered beam is coupled into a single-mode fiber via an objective lens (OL-20x). To accommodate variations in sunlight illumination—particularly under low-intensity conditions such as long-distance measurements, early morning, and late evening—our detection system operates at the single-photon detection level. We placed a series of ND filters to modulate the intensity of sunlight at different times of the day. As shown in part (d) of Fig. \ref{fig.1}, the detection system includes a single-photon counting module (SPCM), which converts each incoming photon from the SMF into an electrical pulse. These pulses are fed into a photon counting module (IDQ), which is interfaced with a computer. The system records the photon arrival events over time and performs a fast Fourier transform (FFT) on the resulting time series to extract the frequency-domain signature associated with the RDE.

Notably, precise optical alignment is critical for the accuracy of the experimental results. The signal beam must be precisely coupled into the single-mode fiber, as any lateral displacement or angular deviation can significantly affect the detection accuracy of the rotational Doppler signal. In multi-wavelength rotational Doppler experiments, the diffraction angles of the SLM vary with wavelength. Thus, each time a filter is replaced, the optical path behind the SLM must be realigned, particularly the fiber coupling system.

\section{Experimental Result}
Firstly, we consider the single wavelength filtered from sunlight by introducing a $532\pm 10~\mathrm{nm}$ filter to test the RDE. Based on our theoretical prediction, although the rotational Doppler signal is profoundly modulated by the coherence length when the spatial coherence is relatively low, one can still observe the RDE with low spatial coherence of a light source. To rigorously confirm that the observed frequency shifts arise from the RDE, we conducted a systematic analysis of frequency shifts across a range of superposition OAM modes. As the inset of Fig.~\ref{fig.2}(a) illustrates, we consider a rotating clover-shaped object whose spiral spectrum is distributed onto $\ell =\pm3,\ell =\pm6,\ell =\pm9$.... Thus, the measured RDE signals exhibit nonzero OAM spectral components in these special modes. We generated and illuminated the rotating object with coherent OAM superposition states ranging from $\ell = 0$ to $\ell = \pm10$. The object was set to rotate at an angular velocity of $\Omega = 4.00 \pm 0.01~\mathrm{r/s}$. As shown in Fig.~\ref{fig.2}(a), we measured $f_3 = 23.90$ Hz,$f_6 = 47.50$ Hz,$f_9= 71.70$ Hz. Compared to the nominal value, the relative errors in the experimentally determined rotation speeds were $\epsilon_{3} = 0.416\%$, $\epsilon_{6} = 1.04\%$, and $\epsilon_{9} = 0.416\%$, all of which fall within the acceptable error margin. Our experimental results provide a clear demonstration that the RDE can be extracted from sunlight.

Considering that the low spatial coherence profoundly influences the intensity of the RDE signal and that the backscattered light from sunlight is usually low,  we conducted further tests by adding a series of ND filters to reduce the intensity of sunlight to photon-level illumination. The rotational speed of the object was set to $\Omega = 8.93 \pm 0.01~\mathrm{r/s}$. The measured frequency spectra with single wavelengths of $\lambda=405~\mathrm{nm},532~\mathrm{nm}, 633~\mathrm{nm}$ are illustrated in Fig. \ref{fig.2}(b), respectively. Due to significant background noise, including ambient light fluctuations and dark counts from the single-photon detector, the Doppler peaks are completely obscured in all three single-wavelength measurements, and no distinguishable RDE signals can be extracted. However, since the rotational Doppler shift is wavelength-independent for a fixed OAM mode and rotation rate, while the noise components are spectrally uncorrelated across wavelengths, we employed a multi-wavelength spectral summation strategy to enhance signal detection. By summing the individual frequency spectra obtained under single-wavelength illumination—namely—denoted as $F_{\text{total}}(\omega) = F_{405}(\omega) + F_{532}(\omega) + F_{635}(\omega)$, we effectively suppressed uncorrelated noise while enhancing the true Doppler signal. The resulting spectrum, shown in Fig. \ref{fig.2}(SUM), exhibits a pronounced and well-defined peak at 53.6 Hz. This frequency corresponds to a rotation rate of approximately $\Omega \approx 8.933\mathrm{r/s}$, which falls within the experimental uncertainty and is in excellent agreement with the theoretical prediction.

\begin{figure}[htp]
    \centering
    \includegraphics[width=1\linewidth]{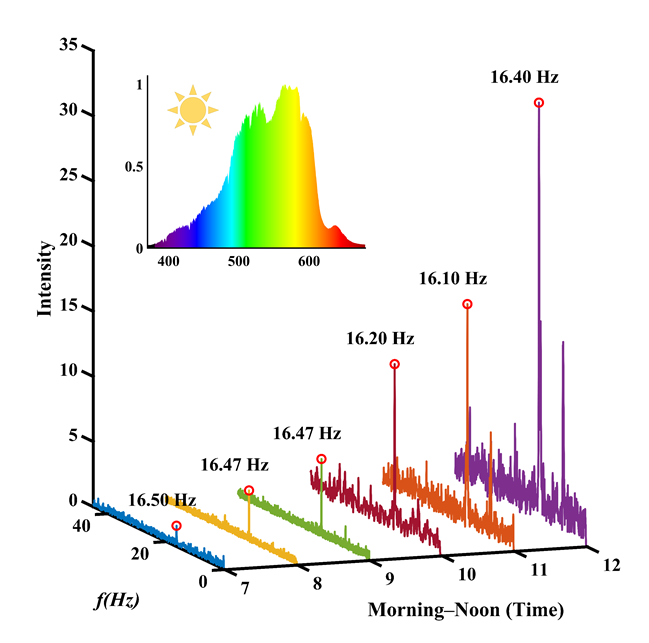}
    \caption{ Sunlight-Based Rotational Doppler Effect: Experimental measurements of frequency shift and signal intensity at different time periods.}
    \label{fig.3}
\end{figure}

To validate the enhancement effect of the multi-wavelength detection strategy demonstrated in the experiments, we conducted an extended study under full-spectrum sunlight illumination. In this test, all bandpass filters were removed from the setup, allowing the system to be directly illuminated by unfiltered sunlight. Given that the preceding experimental results confirmed that combining multiple narrowband signals significantly improves the detectability of rotational Doppler signals under extremely low-light conditions, it is reasonable to expect that, under adequate illumination, the strength of the rotational Doppler signal should increase proportionally with the intensity of broadband solar light. The measured sunlight spectrum using a fiber-optic spectrometer is shown in the inset of Fig.~\ref{fig.3}, which confirms the continuous nature of the solar illumination used in the experiment. All relevant wavelengths contribute simultaneously to the excitation of RDE. In this extended experiment, the RO was driven at a nominal angular velocity of $\Omega = 2.73 \pm 0.01~\mathrm{r/s}$. The measurement campaign began in the early morning when sunlight irradiance was extremely weak and invisible to the naked eye. To ensure stable and precise alignment under such conditions, the entire optical system was pre-aligned and optimized the day before the measurements. At the detection stage, according to the theoretical model, this configuration corresponds to an expected frequency shift of $f \approx 16.38\pm 0.06~\mathrm{Hz}$.
\begin{figure}[htp]
    \centering 
    \includegraphics[width=1\linewidth]{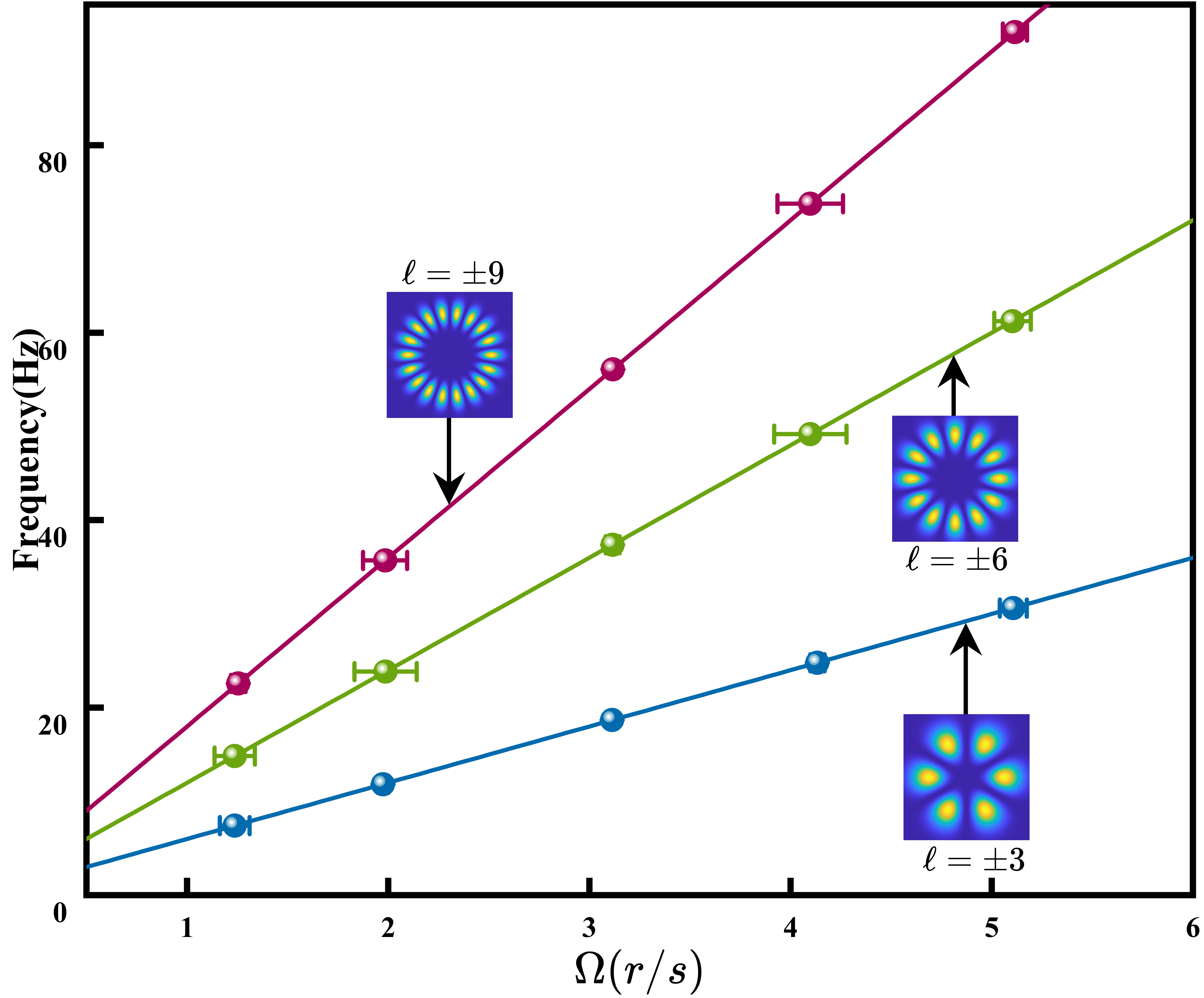}
    \caption{The observed modulation frequencies are plotted as a function of the rotation rate for three different illuminating OAM superposition states, $\ell = \pm3$, $\pm6$, and $\pm9$. The solid lines represent the theoretical predictions based on the rotational Doppler effect, given by $\Delta f \propto\ell \Omega$. Insets display the transverse intensity profiles corresponding to each OAM superposition.}
    \label{fig.4}
\end{figure}
It is worth noting that one inevitable challenge of using broadband sunlight is the chromatic dispersion introduced by the diffraction grating encoded on the SLM. To mitigate this effect and maximize the collection efficiency, the SMF in the original setup was replaced with a multimode fiber (MMF) with a core diameter of $50~\mathrm{\mu m}$. This modification enabled the capture of spatially diverse Doppler-shifted components, though at the cost of admitting some additional stray light. The experimental results are presented in Fig. \ref{fig.3}, where the horizontal axis labeled “Morning–Noon (Time)” indicates the temporal progression during data collection. The rotational Doppler frequency shifts measured at different times of day were consistently $f = (16.4 \pm 0.3)$ Hz. Based on these values, the actual rotation speed of the object was calculated as $\Omega = (2.73 \pm 0.05)\mathrm{r/min}$, with a relative measurement error of $1.4\%$. The measured data reveal a clear correlation between solar illumination intensity and the strength of the observed Doppler signal. As sunlight gradually intensified from early morning to noon, the rotational Doppler signal became more prominent, with the measured peak frequency aligning well with the theoretical prediction. These findings further confirm the effectiveness of the multi-wavelength detection strategy.

To further verify the robustness and general applicability of our proposed method, we conducted a comprehensive statistical analysis of the experimental results on the RDE under sunlight illumination. The analysis examined the relationship between the measured modulation frequency and the rotation rate of the object across different wavelengths and angular velocities. Theoretically, since the rotational Doppler shift is independent of the optical wavelength, the measured data across different wavelengths are expected to exhibit a consistent linear dependence on both the orbital angular momentum $\ell$ and the object's rotation rate $\Omega$, as predicted by the relation $\Delta f \propto\ell \Omega$. We performed measurements using different OAM superposition states, such as $\ell = \pm3$, $\pm6$, and $\pm9$, with five repeated measurements at each rotation rate. Each measurement was performed under a quasi-monochromatic condition, with an optical bandwidth of approximately $20 ~\mathrm{nm}$ centered around the nominal wavelength. The results, along with error bars, are presented in Fig. \ref{fig.4}. The experimental data points closely match the theoretical curves, with measurement uncertainties below $1\%$. The primary sources of uncertainty in our measurements arise from the stability of the rotation stage and slight misalignments in the optical setup during each trial. The agreement between experiment and theory not only confirms the reliability of using sunlight to detect rotational Doppler shifts but also demonstrates that the multi-wavelength detection approach is a robust and effective method for rotational Doppler measurements.

\section{Conclusion}
In conclusion, we have systematically investigated an experimental method for measuring the RDE signal from sunlight. Furthermore, we proposed a multi-wavelength superposition strategy that enables enhanced detection of the RDE under sunlight illumination. The experimental results demonstrate that this approach significantly improves the SNR, allowing clear identification of signals even under low photon count levels. This work opens new possibilities for practical applications of RDE measurements, particularly in passive remote sensing scenarios where rotating targets can be detected using sunlight without any external illumination. Moreover, the multi-wavelength superposition method exhibits remarkable advantages for weak signal detection, improving measurement accuracy in complex, high-background environments. Future work may extend this approach to broader spectral regions—such as infrared or ultraviolet—to evaluate its effectiveness under various illumination conditions.

\section{Acknowledgments}
This work is supported by the Natural Science Foundation of Fujian Province (2023J01007), National Natural Science Foundation of China (12374280, 12034016), the National Key R\&D Program of China (2023YFA1407200), the Fundamental Research Funds for the Central Universities at Xiamen University (20720220030), and the Jiujiang Xun Cheng Talents Program.
%\nocite{*}

%%%%%%%%%%%%%%%%%%%%%%% References %%%%%%%%%%%%%%%%%%%%%%%%%

%%%%%%%%%% If using BibTeX:
\bibliographystyle{apsrev4-2}
\bibliography{apssamp}

\end{document}